\documentclass[10pt]{article}
\usepackage[utf8]{inputenc}
\usepackage[explicit]{titlesec}
\usepackage{geometry}
\usepackage{fancyhdr}
\usepackage{authblk}

\title{More efficient approximation of smoothing splines via space-filling basis selection}

\author{Cheng Meng, Xinlian Zhang, Jingyi Zhang, Wenxuan Zhong, Ping Ma\footnote{cheng.meng25@uga.edu \quad xinlian.zhang25@uga.edu \quad  jingyi.zhang25@uga.edu wenxuan@uga.edu \quad pingma@uga.edu}\\
Department of Statistics, University of Georgia 
}

\date{ }

\usepackage{natbib}
\usepackage{graphicx}
\usepackage{amsmath}
\usepackage{times}
\usepackage{bm}
\usepackage{amsfonts}       
\usepackage[plain,noend]{algorithm2e}
\usepackage{wrapfig}
\usepackage{xcolor}
\usepackage{blindtext}
\usepackage{caption}

\def\H{{\mathcal{H}}}
\def\N{{\mathcal{N}}}
\def\X{{\mathcal{X}}}
\def\S{{\mathcal{S}}}

\newtheorem{thm}{Theorem}
\newtheorem{lem}{Lemma}

\begin{document}

\maketitle
\begin{abstract}
We consider the problem of approximating smoothing spline estimators in a nonparametric regression model. 
When applied to a sample of size $n$, the smoothing spline estimator can be expressed as a linear combination of $n$ basis functions, requiring $O(n^3)$ computational time when the number of predictors $d\geq 2$. 
Such a sizable computational cost hinders the broad applicability of smoothing splines. 
In practice, the full sample smoothing spline estimator can be approximated by an estimator based on  $q$ randomly-selected basis functions, resulting in a computational cost of $O(nq^2)$. 
It is known that these two estimators converge at the identical rate when $q$ is of the order $O\{n^{2/(pr+1)}\}$, where $p\in [1,2]$ depends on the true function $\eta$, and $r > 1$ depends on the type of spline. 
Such $q$ is called the essential number of basis functions.
In this article, we develop a more efficient basis selection method.
By selecting the ones corresponding to roughly equal-spaced observations, the proposed method chooses a set of basis functions with a large diversity. 
The asymptotic analysis shows our proposed smoothing spline estimator can decrease $q$ to roughly $O\{n^{1/(pr+1)}\}$, when $d\leq pr+1$. 
Applications on synthetic and real-world datasets show the proposed method leads to a smaller prediction error compared with other basis selection methods.
\end{abstract}


\section{Introduction}

Consider the nonparametric regression model
$y_i=\eta(x_i)+\epsilon_i (i=1,\ldots,n)$,
where $y_i\in \mathbb{R}$ denotes the $i$th observation of the response, $\eta$ represents an unknown function to be estimated, $x_i \in\mathbb{R}^d$ is the $i$th observation of the predictor variable, and $\{\epsilon_i\}_{i=1}^n$ are independent and identically distributed random errors with zero mean and unknown variance $\sigma^2$.
The function $\eta$ can be estimated by minimizing the penalized least squares criterion,  
\begin{eqnarray}\label{eqn1}
\frac{1}{n}\sum_{i=1}^n\{y_i-\eta(x_i)\}^2+\lambda J(\eta),
\end{eqnarray}
where $J(\eta)$ denotes a quadratic roughness penalty \citep{wahba1990spline,wang2011asymptotics,gu2013smoothing}. 
The smoothing parameter $\lambda$ here administrates the trade-off between the goodness-of-fit of the model and the roughness of the function $\eta$. 
In this paper, expression (\ref{eqn1}) is minimized in a reproducing kernel Hilbert space $\H$, which leads to a smoothing spline estimate for $\eta$.

Although univariate smoothing splines can be computed in $O(n)$ time \citep{reinsch1967smoothing}, it takes $O(n^3)$ time to find the minimizer of (\ref{eqn1}) when $d\geq 2$. 
Such a computational cost hinders the use of smoothing splines for large samples. 
To reduce the computational cost for smoothing splines, extensive efforts have been made to approximate the minimizer of (\ref{eqn1}) by restricting the estimator $\hat{\eta}$ to a subspace $\H_E\subset \H$. 
Let the dimension of the space $\H_E$ be $q$ and the restricted estimator be $\hat{\eta}_E$. 
Compared with the  $O(n^3)$ computational cost of calculating $\hat{\eta}$, the computational cost of $\hat{\eta}_E$ can be reduced to $O(nq^2)$. 
Along this line of thinking, numerous studies have been developed in recent decades.
\citet{luo1997hybrid} and \citet{zhang2004variable} approximated the minimizer of (\ref{eqn1}) using variable selection techniques.
Pseudosplines \citep{hastie1996pseudosplines} and penalized splines \citep{ruppert2009semiparametric} were also developed to approximate smoothing splines. 

Although these methods have already yielded impressive algorithmic benefits, they are usually {\it ad hoc} in choosing the value of $q$.
The value of $q$ regulates the trade-off between the computational time and the prediction accuracy.
One fundamental question is how small $q$ can be in order to ensure the restricted estimator $\hat{\eta}_E$ converge to the true function $\eta$ at the identical rate as the full sample estimator $\hat{\eta}$.
To answer this question, \citet{gu2002penalized,ma2015efficient} developed random sampling methods for selecting the basis functions and established the coherent theory for the convergence of the restricted estimator $\hat{\eta}_E$. 
To ensure that $\hat{\eta}_E$ has the same convergence rate as $\hat{\eta}$, both methods in \citet{gu2002penalized} and \citet{ma2015efficient} require $q$ be of the order $O\{n^{2/(pr+1)+\delta}\}$, where $\delta$ is an arbitrary small positive number, $p\in [1,2]$ depends on the true function $\eta$, and $r$ depends on the fitted spline.
In \citet{gao1999penalized}, it is shown that fewer basis functions are needed to warrant the aforementioned convergence rate if we select the basis functions $\{R(z_j,\cdot)\}_{j=1}^q$, where $\{z_j\}_{j=1}^q$ are roughly equal-spaced.
However, they only provide the theory in the univariate predictor case, and their method cannot be directly applied to multivariate cases.

In this paper, we develop a more efficient computational method to approximate smoothing splines.
The distinguishing feature of the proposed method is that it considers the notion of diversity of the selected basis functions. 
We propose the space-filling basis selection method, which chooses the basis functions with a large diversity, by selecting the ones that correspond to roughly uniformly-distributed observations.
When $d\leq pr+1$, we show that the smoothing spline estimator proposed here has the same convergence rate as the full sample estimator, and the order of the essential number of basis function $q$ is reduced to $O\{n^{(1+\delta)/(pr+1)}\}$.

\section{Smoothing splines and the basis selection method}

Let $\H=\{\eta:J(\eta)<\infty\}$ be a reproducing kernel Hilbert space, where $J(\cdot)$ is a squared semi-norm. 
Let $\N_J=\{\eta:J(\eta)=0\}$ be the null space of $J(\eta)$ and assume that $\N_J$ is a finite-dimensional linear subspace of $\H$ with basis $\{\xi_i\}_{i=1}^m$ in which $m$ is the dimension of $\N_J$.
Let $\H_J$ be the orthogonal complement of $\N_J$ in $\H$ such that $\H=\N_J\oplus \H_J$.  The space $\H_J$ is a reproducing kernel Hilbert space with  $J(\cdot)$ as the squared norm. 
The reproducing kernel of $\H_J$ is denoted by $R_J(\cdot,\cdot)$.
The well-known representer theorem \citep{wahba1990spline} states that there exist vectors $d=(d_1,\ldots,d_m)^T\in\mathbb{R}^m$ and $c=(c_1,\ldots,c_n)^T\in\mathbb{R}^n$, such that the minimizer of (\ref{eqn1}) in $\H$ is given by
$\eta(x)=\sum_{k=1}^m d_k\xi_k(x)+\sum_{i=1}^n c_i R_J(x_i,x).$
Let $Y=(y_1,\ldots,y_n)^T$ be the vector of response observations,
$S$ be the $n\times m$ matrix with the $(i,j)$th entry  $\xi_j(x_i)$, and $R$ be the $n\times n$ matrix with the $(i,j)$th entry $R_J(x_i,x_j)$.
Solving the minimization problem in (\ref{eqn1}) thus is equivalent to solving 

\begin{equation}\label{eqn4}
(\hat{d}, \hat{c})=
\underset{d,c}{\mathrm{argmin}}\frac{1}{n}(Y-Sd-Rc)^T(Y-Sd-Rc)+\lambda c^TRc,
\end{equation} 
where the smoothing parameter $\lambda$ can be selected based on the generalized cross-validation criterion \citep{wahba1978smoothing}.
In a general case where $n\gg m$ and $d\geq 2$, the computation cost for calculating $(\widehat{d}, \widehat{c})$ in equation (\ref{eqn4}) is of the order $O(n^3)$, which is prohibitive when the sample size $n$ is considerable. 
To reduce this computational burden, one can restrict the full sample estimator $\hat{\eta}$ to a subspace $\H_E\subset \H$, where $\H_E=\N_J\oplus \mbox{span}\{R_J(x_i^*,\cdot),i=1,\ldots,q\}$.
Here, $\H_E$, termed as the effective model space, can be constructed by selecting a subsample $\{x_i^*\}_{i=1}^q$ from $\{x_i\}_{i=1}^n$.
Such an approach is thus called the basis selection method.

Denote a matrix $R_*\in \mathbb{R}^{n\times q}$ with the $(i,j)$th entry $R_J(x_i,x_j^*)$  and $R_{**} \in \mathbb{R}^{q\times q}$ with the $(i,j)$th entry $R_J(x_i^*,x_j^*)$. 
The minimizer of (\ref{eqn1}) in the effective model space $\H_E$ thus can be written as
$\eta_E(x)=\sum_{k=1}^m d_k\xi_k(x)+\sum_{i=1}^q c_i R(x^*_i,x)$ and the coefficients, $d_E=(d_1,\ldots,d_m)^T$ and $c_E=(c_1,\ldots,c_q)^T$ can be obtained through solving
\begin{eqnarray}
(\hat{d}_E, \hat{c}_E)=
\underset{d_E,c_E}{\mathrm{argmin}}\frac{1}{n}(Y-Sd_E-R_*c_E)^T(Y-Sd_E-R_*c_E)+\lambda c_E^TR_{**}c_E.
\end{eqnarray}

The evaluation of the restricted estimator $\hat{\eta}_E$ at sample points thus satisfies
$\hat{\eta}_E=S\hat{d}_E+R_*\hat{c}_E$,
where $\hat{\eta}_E=\{\hat{\eta}_E(x_1),\ldots,\hat{\eta}_E(x_n)\}^T$.
In a general case where $m\ll q\ll n$, the overall computational cost for the basis selection method is $O(nq^2)$, which is a significant reduction compared with $O(n^3)$. 
Recall that the value of $q$ controls the trade-off between the computational time and the prediction accuracy. 
To ensure that $\hat{\eta}_E$ converges to the true function $\eta$ at the same rate as $\hat{\eta}$, researchers showed that the essential number of basis functions $q$ needs to be of the order $O\{n^{2/(pr+1)+\delta}\}$, where $\delta$ is an arbitrary small positive number \citep{kim2004smoothing,ma2015efficient}. 
In the next section, we present the proposed space-filling basis selection method, which reduces such an order to $O\{n^{(1+\delta)/(pr+1)}\}$.

\section{Space-filling basis selection}
\subsection{Motivation and Notations}
To motivate the development of the proposed method, we first re-examine the ensemble learning methods, which are well-known in statistics and machine learning \citep{dietterich2002ensemble,rokach2010ensemble}.
To achieve better predictive performance than a single learner \footnote{A learner is either a model or a learning algorithm.}, ensemble learning methods first build a committee which consists of a number of different learners, then aggregate the predictions of these learners in the committee.
The aggregation is usually achieved by employing the majority vote or by calculating a weighted average. 
The diversity among the learners in the committee holds the key to the success of the ensemble learning methods. 
A large diversity in the committee yields a better performance of ensemble learning methods \citep{kuncheva2003measures}.

The restricted smoothing spline estimator $\hat{\eta}_E$ can be considered as an ensemble learning method. 
In particular, the prediction of $\hat{\eta}_E$ is conducted by taking a weighted average of the predictions 
of the selected basis functions $R_J(x_i^*,\cdot)$, $i\in\{1,\ldots,q\}$, in addition to the basis functions in the null space $\N_J$. 
Inspired by \citet{kuncheva2003measures}, we propose to select a subsample $\{x_i^*\}_{i=1}^q$, such that the diversity among the basis functions $\{R_J(x_i^*,\cdot)\}_{i=1}^q$ is as large as possible.
One crucial question is how to measure the diversity among a set of basis functions.
Notice that adjacent data points, i.e., $x_i^*\approx x_j^*$ ($i,j\in\{1,\ldots,q\}$) yields similar basis functions, i.e.,  $R_J(x_i^*,\cdot)\approx R_J(x_j^*,\cdot)$. 
On the other hand, if $x_i^*$ is far away from $x_j^*$, the basis function $R_J(x_i^*,\cdot)$ tends to be different from $R_J(x_j^*,\cdot)$.
These observations inspire us to select a set of basis functions $\{R_J(x_i^*,\cdot)\}_{i=1}^q$ where $\{x_i^*\}_{i=1}^q$ are as uniformly-distributed as possible. 
The uniformly-distributed property, usually termed as the space-filling property in the experimental design literature \citep{pukelsheim2006optimal}, can be systematically measured by the star discrepancy.

Since the star discrepancy is defined for data in $[0,1]^d$, in practice, we need to map the data with arbitrary distribution to this domain. 
Suppose $\X_n=\{x_i\}_{i=1}^n$ are independent and identically distributed observations generated from the cumulative distribution function $F$ with bounded support $\mathcal{D}\subset\mathbb{R}^d$. 
Suppose $\tau$ is a transformation, such that $\{\tau(x_i)\}_{i=1}^n$ has the uniform distribution on $[0,1]^d$. In a simple case where $d=1$ and $F$ is known, we can find the transformation $\tau$ by setting $\tau=F$.
In a more general case where $d>1$ and $F$ is unknown, the transformation $\tau$ can be calculated via the optimal transport theory \citep{villani2008optimal}.
However, finding the exact solution via the optimal transport theory can be time-consuming. 
Instead, 
one may approximate the transformation $\tau$ using the iterative transformation approach \citep{pukelsheim2006optimal} or the sliced optimal transport map approach \citep{rabin2011wasserstein}.
The computational cost of these two approaches is of the order $O\{Kn\log(n)\}$, where $K$ denotes the number of iterations \citep{kolouri2018sliced,cuturi2014fast,bonneel2015sliced}.
Such a computational cost is negligible compared with the computational cost of the proposed method. 
In practice, the data can always be preprocessed using the transformation of $\tau$. 
Without loss of generality, $\X_n$ may be assumed to be independent and identically distributed observations generated from the uniform distribution on $[0,1]^d$.

\subsection{Star discrepancy and space-filling design}

Let $a=(a_1,\ldots,a_d)^T\in[0,1]^d$, $[0, a)=\prod_{j=1}^d[0,a_j)$ be a hyper-rectangle and $1\{\cdot\}$ be the indicator function. The local discrepancy and the star discrepancy are defined as follows \citep{fang2005design, pukelsheim2006optimal}. 

Given $\mathcal{X}_q=\{x_1,\ldots,x_q\}$ in $[0,1]^d$ and a hyper-rectangle $[0,a)$, the corresponding local discrepancy is defined as $D(\mathcal{X}_q,a)=|\frac{1}{q}\sum^q_{i=1}1\{x_i\in[0,a)\}-\prod^d_{j=1}a_j|.$
The star discrepancy corresponding to $\mathcal{X}_q$ is defined as
$D^*(\X_q)=\sup_{a\in [0,1]^d}D(\X_q,a).$

\begin{figure}[ht]
    \begin{center}
        \begin{tabular}{cc}
            \includegraphics[scale = .15]{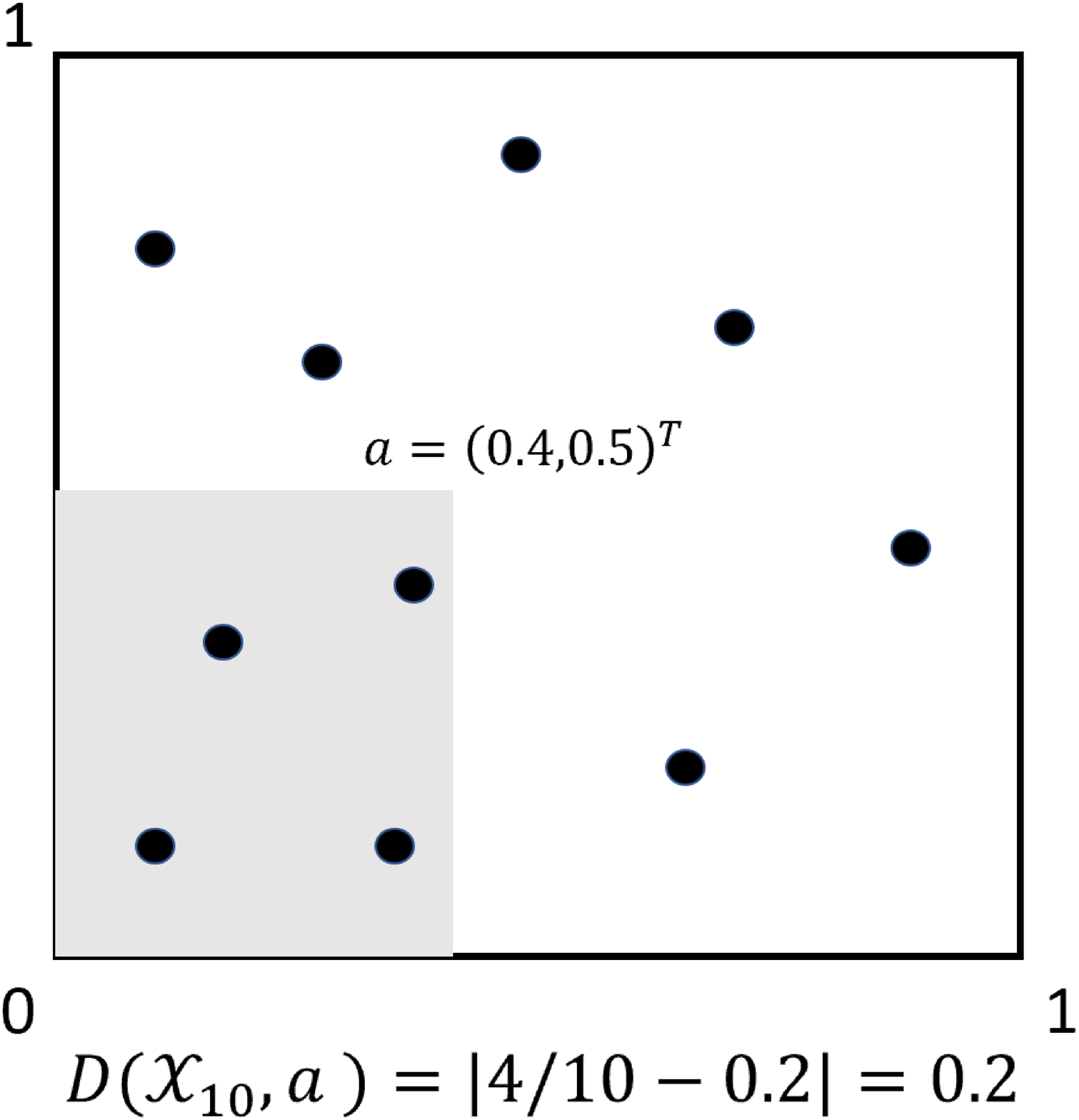}  & \includegraphics[scale = .475]{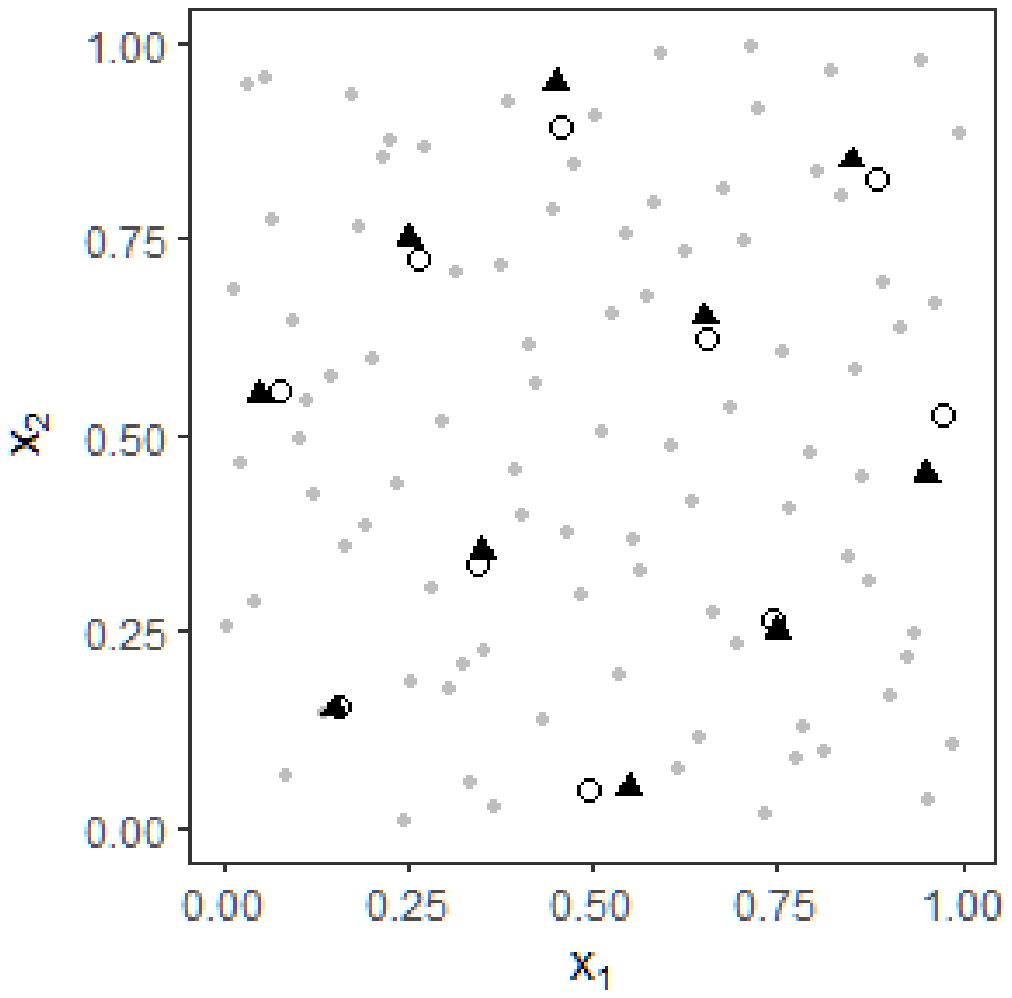} \\
        \end{tabular}
        \caption{Left panel: A toy example for local discrepancy. A set of ten data points are generated in $[0,1]^2$, and four of them locate in the rectangular $[0,a)$, where $a=(0.4,0.5)^T$. The local discrepancy is $|4/10-0.4\times 0.5|=0.2$. Right panel: An illustration for the proposed basis selection method. The data points are labelled as gray dots. The nearest neighbor data point for each design point (black triangle) is labeled as a circle.}\label{fig1}
    \end{center}
\end{figure}

The local discrepancy $D(\X_q, a)$ measures the difference between the volume of the hyper-rectangle $[0, a)$ and the fraction of the data points located in $[0, a)$.
The local discrepancy is illustrated in the left panel of Fig. \ref{fig1}. 
The star discrepancy $D^*(\X_q)$ calculates the supreme of all the local discrepancy over $a\in [0,1]^d$. 
In other words, the smaller the $D^*(\X_q)$ is, the more space-filling the data points $\X_q$ are \citep{fang2005design}.

\citet{chung1949estimate} showed that when $\X_q$ is generated from the uniform distribution in $[0,1]^d$, $D^*(\X_q)$ converges to 0 with the order of convergence $O[\{\log\log(q)/q\}^{1/2}]$. 
Faster convergence rate can be achieved using space-filling design methods \citep{pukelsheim2006optimal} and the low-discrepancy sequence method \citep{halton1960efficiency,soboldistribution,owen2003quasi}. 
The space-filling design methods, developed in the experimental design literature, aim to generate a set of design points that can cover the space as uniformly as possible. For further details, please refer to \citet{wu2011experiments,fang2005design,pukelsheim2006optimal}. 
The low-discrepancy sequence method 
Such a method is frequently applied in the field of quasi-Monte Carlo and is extensively employed for numerical integration. 
For a Sobol sequence $\S_q$, one type of low-discrepancy sequence, it is known that $D^*(\S_q)$ is of the order $O\{\log(q)^d/q\}$, which is roughly the square order $D^*(\X_q)$ for fixed $d$. For more in-depth discussions on the quasi-Monte Carlo methods, see  e.g., \cite{lemieux2009book,leobacher2014introduction,dick2013high} or Chapter 5 in \cite{,glasserman2013monte} and references therein.

Existing studies suggested that space-filling property can be exploited to achieve a fast convergence rate for numerical integration and response surface estimation \citep{fang2005design,pukelsheim2006optimal}. 
These results inspire us to select the space-filling basis functions in smoothing splines. 
Unfortunately, the existing techniques of space-filling design cannot be applied to our basis selection problem directly due to the following fact.
The design space in space-filling design methods is usually
continuous, whereas our sample space $\{x_i\}_{i=1}^n$ is finite and discrete.
We propose an algorithm to overcome the barrier.

\subsection{Main algorithm}

We shall develop a space-filling basis selection method, in which we select the space-filling data points in a computationally-attractive manner. 
First, a set of design points $\S_q=\{s_i\}_{i=1}^q \in [0,1]^d$ are generated, either using low-discrepancy sequence or space-filling design methods.
Subsequently, the nearest neighbor $x^*_i$ is selected for each $s_i$, from the sample points $\{x_i\}_{i=1}^n$. Thus, $\{x_i^*\}_{i=1}^q$ can approximate the design points $\S_q$ well, if $x^*_i$ is sufficiently close to $s_i$, for $i=1,\ldots q$. The proposed method is summarized as follows. 
\begin{itemize}
\item \textit{Step} 1. Generate a set of design points $\{s_i\}_{i=1}^q$ from $[0,1]^d$.
\item \textit{Step} 2. Select the nearest neighbor for each design point $s_i$ from $\{x_i\}_{i=1}^n$. Let the selected data points be $\{x^*_i\}_{i=1}^q$,.
\item \textit{Step} 3. Minimize the penalized least squares criterion (\ref{eqn1}) over the following effective model space
$\H_E=\mathcal{N}_J \oplus \mbox{span}\{R_J(x_i^*,\cdot), i=1,\ldots,q\}$
\end{itemize}

The proposed algorithm is illustrated through a toy example in the right panel of Fig. 1. One hundred data points (gray dots) are generated from the uniform distribution in $[0,1]^2$, and a set of design points (black triangles) are generated through the max projection design \citep{joseph2015maximum}, a recently developed space-filling design method. The nearest neighbor to each design point is selected (circle). It is observed that the selected subsample is space-filling since it can effectively approximate the design points. 

\begin{figure}[ht]
        \includegraphics[scale = .4]{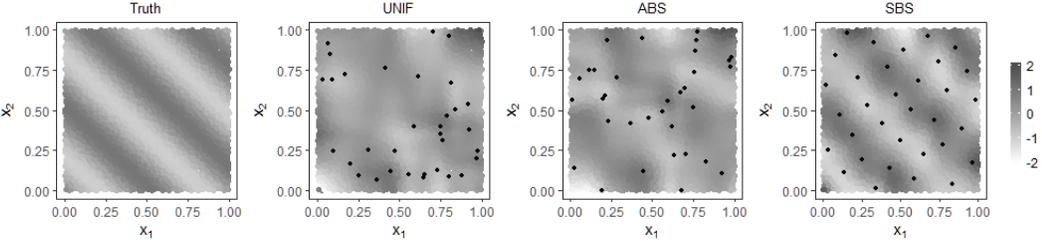} 
        \caption{Comparison of different basis selection methods. The leftmost panel shows the heat map for the true function. The heat maps for the spline estimates based on the uniform basis selection method (UNIF), the adaptive basis selection method (ABS), and the proposed space-filling basis selection method (SBS) are presented in the other three panels, respectively. Black dots are the sampled basis functions. We observe that the proposed method outperforms the other methods in approximating the true function.} \label{fig2}
\end{figure}
In Fig. \ref{fig2}, the proposed space-filling basis selection method is compared with the uniform basis selection method \citep{gu2002penalized} and the adaptive basis selection method \citep{ma2015efficient} using a two-dimensional toy example. 
We generate 5000 data points from the uniform distribution in $[0,1]^2$. 
The leftmost panel in Fig. \ref{fig2} presents the heat map for the true response surface $y=\sin\{5(x_1+x_2)\}$. 
The dimension of the effective model space $q$ is set to be $5\times(5000)^{2/9}\approx 33$, for all basis selection methods. 
The selected basis functions are labeled as solid dots in each panel. The right three panels of Fig.~\ref{fig2} plot the heat maps of the spline estimates of all three basis selection methods.
In the uniform basis selection method, the default random number generator in R is employed to select the basis functions.
It is observed that the selected points are not uniformly distributed.
This is a very common phenomenon for uniform basis selection since the randomly selected points do not necessarily look uniformly-distributed, especially when the number of selected points is small.
In contrast, it is observed that the basis functions selected by the proposed method are space-filling. 
Using the space-filling design techniques, the proposed method overcomes the pitfall of uniform basis selection method and uniformly-distribute the selected points. 
The true response can be better estimated using the proposed method than using other methods.

Now we calculate the computational cost of the proposed method. In Step 1, the design points can be generated beforehand; thus, the computational cost in Step 1 can be ignored.
For the nearest neighbor search in Step 2, we employ the $k$-d tree method, which takes $O\{n\log(n)\}$ flops \citep{bentley1975multidimensional,wald2006building}. The computational cost of this step can be further reduced if we are willing to sacrifice the accuracy of the searching results, e.g., using those approximate nearest neighbor searching algorithms \citep{altman1992introduction,arya1994optimal}. 
For Step 3, computing the smoothing spline estimates in the restricted space $\H_E$ is of the order $O(nq^2)$, as discussed in Section 2.2. In summary, the overall computational cost for the space-filling basis selection method is of the order $O(nq^2)$.
\section{Convergence rates for function estimation}
Recall that the data points are assumed to be generated from the uniform distribution on $[0,1]^d$. Thus, for each coordinate $x$, the corresponding marginal density $f_X(\cdot)=1$. We define that $V(g)=\int_{[0,1]^d} g^2\mbox{d}x$. 
The following four regularity conditions are required for the asymptotic analysis, and the first three are the identical conditions employed by \cite{ma2015efficient}, in which one can find more technical discussions.

Condition 1. The function $V$ is completely continuous with respect to $J$; 

Condition 2. for some $\beta>0$ and $r>1$, $\rho_\nu>\beta\nu^r$ for sufficiently large $\nu$;

Condition 3. for all $\mu$ and $\nu$, $\mbox{var}\{\phi_\nu(x)\phi_\mu(x)\}\le C_1$, where $\phi_\nu$, $\phi_\mu$ are the eigenfunctions associated with $V$ and $J$ in $\H$, $C_1$ denotes a positive constant; 

Condition 4. for all $\mu$ and $\nu$, $\mathcal{V}(g_{\nu,\mu})\le C_2$, where $\mathcal{V}(\cdot)$ denotes the total variation, $g_{\nu,\mu}(x)=\phi_\nu(x)\phi_\mu(x)$, and $C_2$ represents a positive constant.
The total variation here is defined in the sense of Hardy and Krause \citep{owen2003quasi}. As a specific case when $d=1$, the total variation $\mathcal{V}(g)=\int|g'(x)|\mbox{d}x$, revealing that a smooth function displays a small total variation.
Intuitively, the total variation measures how wiggly the function $g$ is.


Condition 1 indicates that there exist a sequence of eigenfunctions $\phi_\nu\in\H$  and the associated sequence of eigenvalues $\rho_\nu \uparrow \infty$ satisfying $V(\phi_\nu,\phi_\mu)=\delta_{\nu\mu}$ and $J(\phi_\nu,\phi_\mu)=\rho_\nu\delta_{\nu\mu}$,  
where $\delta_{\nu\mu}$ is the Kronecker delta. The growth rate of the eigenvalues $\rho_\nu$ dictates how fast $\lambda$ should approach to 0, and further what the convergence rate of smoothing spline estimates is~\citep{gu2013smoothing}. 
Notice that the eigenfunctions $\phi_\nu$s have a close relationship with the Demmler-Reinsch basis, which are  orthogonal vectors representing $l_2$ norm ~\citep{ruppert2002selecting}.
The eigenfunctions $\phi_\nu$s can be calculated explicitly in several specific scenarios.
For instance, $\phi_\nu$s are the sine and cosine functions when $J(\eta)=\int_0^1 (\eta^{''})^2\mbox{d}x$, where $\eta$ denotes a periodic function on $[0,1]$.
For more details on the construction of $\phi_\nu$s can be found in Section 9.1 of \citet{gu2013smoothing}.  


We now present our main theoretical results, and all the proofs are relegated to the Supplementary Material.
For a set of design points $\S_q$ of size $q$, we now assume the star discrepancy $D^*(\S_q)$ converges to zero at the rate of $O\{\log(q)^d/q\}$, or $O\{q^{-(1-\delta)}\}$ for an arbitrary small positive number $\delta$.
Such a convergence rate is warranted if $\S_q$ is generated from a low-discrepancy sequence or space-filling design methods, as discussed in Section 3.1.
Recall that the proposed method aims to select a subsample that is space-filling, and the key to success depends on whether the chosen subsample $\X_q^*$ can effectively approximate the design points $\S_q$.
The following lemma bounds the difference between $\X_q^*$ and $\S_q$ in terms of star discrepancy.
\begin{lem}\label{lem4.1} 
Suppose $d$ is fixed and $D^*(\S_q)=O\{q^{-(1-\delta)}\}$, for any arbitrary small $\delta>0$. If $q=O(n^{1/d})$, as $n \rightarrow \infty$, we have $D^*(\X^*_q)=O_p\{q^{-(1-\delta)}\}.$
\end{lem}

Lemma \ref{lem4.1} states that the selected subsample $\X_q^*$ can effectively approximate the design points $\S_q$ in the sense that the convergence rate of $D^*(\X^*_q)$ is similar to that of $D^*(\S_q)$. 
The following theorem is the Koksma--Hlawka inequality, which will be used in proving our main theorem. See \citet{kuipers2012uniform} for a proof.
\begin{thm}[Koksma--Hlawka inequality]\label{thm4.2}
Let $\mathcal{T}_q=\{t_1,\ldots,t_q\}$ be a set of data points in $[0,1]^d$,  and $h$ be a function defined on $[0,1]^d$ with bounded total variation $\mathcal{V}(h)$.  We have
$|\int_{[0,1]^d}h(x)\mbox{d}x-\sum_{i=1}^qh(t_i)/q|\leq D^*(\mathcal{T}_q)\mathcal{V}(h).$
\end{thm}

Combining Lemma \ref{lem4.1} and Theorem \ref{thm4.2} and set $h=g_{\nu,\mu}$, $\mathcal{T}_q=\X_q^*$ yields the following lemma.

\begin{lem}\label{lem3}
If $q=O(n^{1/d})$, under Condition 4, for all $\mu$ and $\nu$, we have 
\begin{eqnarray*}
\left|\int_{[0,1]^d}\phi_\nu\phi_\mu \mbox{d}x -\frac{1}{q}\sum_{j=1}^q\phi_\nu(x_j^*)\phi_\mu(x_j^*) \right|=O_p\{q^{-(1-\delta)}\}.
\end{eqnarray*}
\end{lem}

Lemma \ref{lem3} shows the advantage of $\{x_i^*\}_{i=1}^q$, the subsample selected by the proposed method, over a randomly selected subsample $\{x_i^+\}_{i=1}^q$. To be specific, as a direct consequence of Condition 3, we have 
$E[\int_{[0,1]^d}\phi_\nu\phi_\mu \mbox{d}x -\sum_{j=1}^q\phi_\nu(x_j^+)\phi_\mu(x_j^+)/q]^2=O(q^{-1}),$
for all $\mu$ and $\nu$.
Lemma 2 therefore implies the subsample $\X_q^*$ can more efficiently approximate the integration $\int_{[0,1]^d}\phi_\nu\phi_\mu \mbox{d}x$, for all $\mu$ and $\nu$. 
We now present our main theoretical result.


\begin{thm}\label{thm3}
Suppose $\sum_i\rho_i^pV(\eta_0,\phi_i)^2<\infty$ for some $p\in[1,2]$ and $\delta$ is an arbitrary small positive number. Under Conditions 1 - 4,  $q=O(n^{1/d})$,  as $\lambda \rightarrow 0$ and $q^{1-\delta}\lambda^{1/r} \rightarrow \infty$,  we have $(V+\lambda J)(\hat{\eta}_E-\eta_0)=O_p(n^{-1}\lambda^{-1/r}+\lambda^p)$. In particular, if $\lambda \asymp n^{-r/(pr+1)}$, the estimator achieves the optimal convergence rate
$(V+\lambda J)(\hat{\eta}_E-\eta_0)=O_p\{n^{-pr/(pr+1)}\}.$
\end{thm}

It is shown in Theorem 9.17 of \citet{gu2013smoothing} that the full sample smoothing spline estimator $\hat{\eta}$ has the convergence rate, 
$(V+\lambda J)(\hat{\eta}-\eta_0)=O_p(n^{-pr/(pr+1)})$ under some regularity conditions.
Theorem \ref{thm3} thus states that proposed estimator $\hat{\eta}_E$ achieves the same convergence rate as the full sample estimator $\hat{\eta}$, under two extra conditions imposed on $q$ 
(a) $q=O(n^{1/d})$, and (b) $q^{1-\delta}\lambda^{1/r} \rightarrow \infty$ as $\lambda \rightarrow 0$.
Moreover, Theorem \ref{thm3} indicates that to achieve the identical convergence rate as the full sample estimator $\hat{\eta}$, the proposed approach requires a much smaller number of basis functions, in the case when $\lambda \asymp n^{-r/(pr+1)}$.
$q^{1-\delta}\lambda^{1/r} \rightarrow \infty$  indicates an essential choice of $q$ for the proposed estimator should satisfy $q=O\{n^{(1+\delta)/(pr+1)}\}$, when $\lambda \asymp n^{-r/(pr+1)}$.
For comparison, both the random basis selection \citep{gu2002penalized} and the adaptive basis selection method \citep{ma2015efficient} require the essential number of basis functions to be $q=O\{n^{2/(pr+1)+\delta}\}$. 
As a result, the proposed estimator is more efficient since it reduces the order of the essential number of basis functions.

Given $q=O(n^{1/d})$, when $d\leq pr+1$, it follows naturally when $q^{1-\delta}\lambda^{1/r} \rightarrow \infty$  is satisfied. Otherwise, when $d>pr+1$, $q=O(n^{1/d})$ becomes sufficient but not necessary for $q^{1-\delta}\lambda^{1/r} \rightarrow \infty$. We thus stress that the essential number of basis functions for the proposed method, $q=O\{n^{(1+\delta)/(pr+1)}\}$, can only be achieved when $d\leq pr+1$.
The parameter $p$ in Theorem 2 is closely associated with the true function $\eta_0$ and will affect the convergence rate of the proposed estimator. 
Intuitively, the larger the $p$ is, the smoother the function $\eta_0$ will be.
For $p\in[1,2]$, the optimal convergence rate of $(V+\lambda J)(\hat{\eta}_E-\eta_0)$ falls in the interval $[O_p(n^{-r/(r+1)}), O_p(n^{-2r/(2r+1)})]$.
To the best of our knowledge, the problem of selecting the optimal $p$ has rarely been studied, and one exception is \citet{serra2017adaptive}, where the author studied such a problem in one-dimensional cases.
\citet{serra2017adaptive} provided a Bayesian approach for selecting the optimal parameter, named $\beta$, which is known to be proportional to $p$.
Nevertheless, since the constant $\beta/p$ is usually unknown, such an approach still cannot be used to select the optimal $p$ in practice. Furthermore, whether such an approach can be extended to the high-dimensional cases remains unclear.

For the dimension of the effective model space $q$, a suitable choice is $q = n^{(1+\delta)/(4p+1)+\delta}$ in the following two cases. Case 1. Univariate cubic smoothing splines with the penalty $J(\eta) = \int_0^1(\eta'')^2$, $r=4$ and $\lambda \asymp n^{-4/(4p+1)}$; Case 2. Tensor-product splines with $r=4-\delta^*$, where $\delta^*>0$.
For $p \in [1,2]$, the dimension roughly lies in the interval $(O(n^{1/9}), O(n^{1/5}))$.

\section{Simulation Results}
To assess the performance of the proposed space-filling basis selection method, we carry out extensive analysis on simulated datasets. We report some of them in this section.
We compare the proposed method with uniform basis selection and adaptive basis selection, and report both prediction error and running time.

The following four functions on $[0, 1]$  \citep{lin2006component} are used as the building blocks in our simulation study, 
$g_1(t)=t$, $g_2(t)=(2t-1)^2$, $g_3(t)=\sin(2\pi t)/\{2-\sin(2\pi t)\}$, and $g_4(t)=0.1\sin(2\pi t)+0.2\cos(2\pi t)+0.3\sin(2 \pi t)^2+0.4\cos(2 \pi t)^3+0.5\sin(2 \pi t)^3$. In addition, we also use the following functions on $[0,1]^2$ \citep{wood2003thin} as the building blocks,
\begin{align*}
h_1(t_1,t_2)=\{0.75/(\pi \sigma_1\sigma_2)\}\times\exp\{-(t_1-0.2)^2/\sigma_1^2 - (t_2-0.3)^2/\sigma_2^2\}, \\
h_2(t_1,t_2)=\{0.45/(\pi\sigma_1\sigma_2)\}\times\exp\{-(t_1-0.7)^2/\sigma_1^2 - (t_2-0.8)^2/\sigma_2^2\},
\end{align*}
where $\sigma_1=0.3$ and $\sigma_2=0.4$. The signal-to-noise ratio (SNR), defined as $\mbox{var}\{\eta(X)\}/\sigma^2$, is set to be at two levels: 5 and 2. We generate replicated samples with sample sizes $n=\{2^{10},2^{11},\ldots,2^{14}\}$ and dimensions $d=\{2,4,6\}$ uniformly on $[0,1]^p$ from the following four regression settings,

(1) A 2-d function $g_1(x_1x_2)+g_2(x_2)+g_3(x_1)+g_4(x_2)+g_3\{(x_1+x_2)/2\}$; 

(2) A 2-d function $h_1(x_1,x_2)+h_2(x_1,x_2)$; 

(3) A 4-d function $g_1(x_1)+g_2(x_2)+g_3(x_3)+2g_1\{(x_1+x_4)/2\}+2g_2\{(x_2+x_3)/2\}+2g_3\{(x_1+x_3)/2\}$; 

(4) A 6-d function $h(x_1,x_2)+h(x_1,x_5)$.

In the simulation, $q$ is set to be $5n^{2/9}$ and $10n^{1/9}$, based on the asymptotic results. 
To combat the curse of dimensionality, we fit smoothing spline analysis of variance models with all main effects and two-way interactions.
The prediction error is measured by the mean squared error (MSE), defined as $[\sum_{i=1}^{n_0}\{\hat{\eta}_E(t_i)-\eta_0(t_i)\}^2]/n_0$, where $\{t_i\}_{i=1}^{n_0}$ denotes an independent testing dataset uniformly generated on $[0,1]^p$ with $n_0=5000$. 
The max projection design \citep{joseph2015maximum} is used to generate design points in Step 1 of the proposed method.
Our empirical studies suggest that the Sobol sequence and other space-filling techniques, e.g., the Latin hypercube design \citep{pukelsheim2006optimal} and the uniform design \citep{fang2000uniform}, also yield similar performance. 

\begin{figure}[ht]
    \begin{center}
        \begin{tabular}{c}
            \includegraphics[scale = .31]{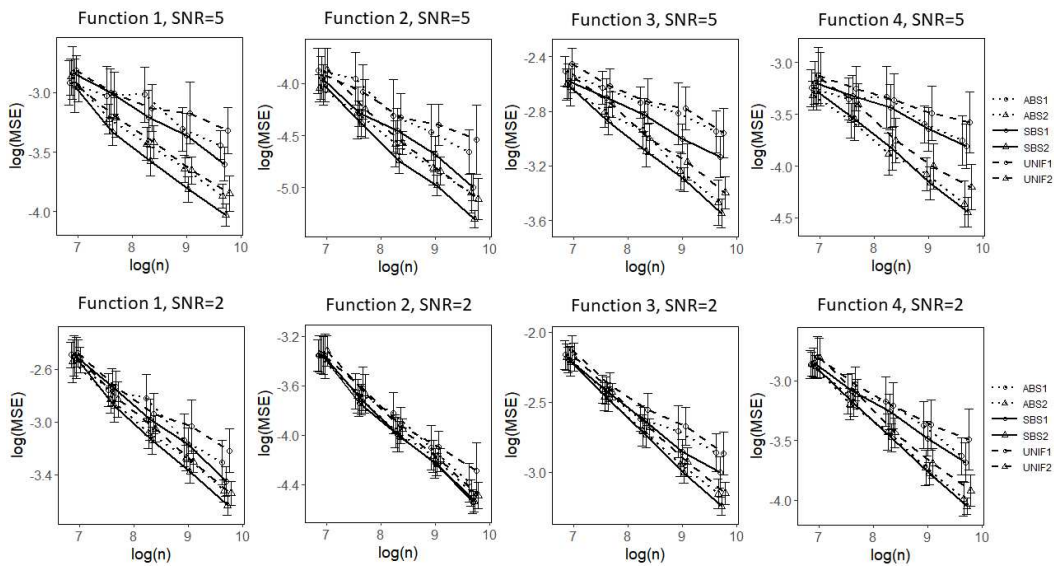}
        \end{tabular}
        \caption{Simulation under different settings (from left to right) with SNR being five (the upper row) and two (the lower row). The prediction errors are plotted versus sample size $n$
        The vertical bars are standard error bars obtained from 20 replicates. The solid lines, dotted lines, and dashed lines refer to the space-filling basis selection (SBS), the adaptive basis selection (ABS), and the uniform basis selection (UNIF), respectively. The lines with triangles and those with circles represent $q=5n^{2/9}$ and $q=10n^{1/9}$, respectively.}\label{plot3}
    \end{center}
\end{figure}

Figure \ref{plot3} shows the MSE against the sample size on the log-log scale. Each column presents the results of a function setting as described above. We set the signal-to-noise ratio to be five and two in the upper row and the lower row, respectively. 
We use solid lines for the proposed method, dotted lines for adaptive basis selection method, and dashed lines for uniform basis selection method.
The number of the basis functions $q$ is $5n^{2/9}$ and $10n^{1/9}$ for the lines with triangles and the lines with circles, respectively. 
The vertical bars represent standard error bars obtained from 20 replicates. The full sample estimator is omitted here due to the high computation cost.
It is observed that the space-filling basis selection method provides more accurate smoothing spline predictions than the other two methods in almost all settings.
Notably, the lines with circles for the space-filling basis selection method displays a linear trend similar to the lines with triangles for the other two methods. 
This observation reveals the proposed estimator yields a faster convergence rate than the other two methods. 

More simulation results can be found in the Supplementary Material, in which we consider the regression functions that exhibit several sharp peaks. 
In those cases, the results suggest that both the space-filling basis selection method and the adaptive basis selection method outperform the uniform basis selection, whereas neither the space-filling basis selection method nor the adaptive basis selection method dominates each other.
Moreover, the proposed space-filling basis selection method outperforms the adaptive basis selection method as the sample size $n$ gets larger.

\begin{table}[ht]
   \label{table1}
   \centering
   \caption{Means and standard errors (in parentheses) of the computational time, in CPU seconds, for multivariate cases, based on 20 replicates.}
\begin{tabular}{ccccc}
True function & SNR & UNIF         & ABS          & SBS          \\
Function 1    & 5   & 0.97(0.15)   & 0.90(0.05)   & 0.90(0.04)   \\
              & 2   & 0.92(0.10)   & 0.87(0.04)   & 0.87(0.06)   \\
Function 2    & 5   & 0.88(0.04)   & 0.87(0.03)   & 0.90(0.06)   \\
              & 2   & 0.86(0.05)   & 0.85(0.02)   & 0.90(0.06)   \\
Function 3    & 5   & 3.92(0.24)   & 3.95(0.24)   & 4.04(0.19)  \\
              & 2   & 4.08(0.30)   & 4.51(0.66)   & 4.27(0.39)  \\
Function 4    & 5   & 12.95(0.61)  & 15.10(3.20)  & 15.45(3.04) \\
              & 2   & 14.33(1.44)  & 13.72(1.02)  & 14.25(1.09)
\end{tabular}
\end{table}
Table 1 summarizes the computing times of model-fitting using all  methods on a synthetic dataset with $n=2^{14}$ and $q=5n^{2/9}$. The simulation is replicated for 20 runs using a computer with an Intel 2.6 GHz processor. 
In Table 1, UNIF, ABS, and SBS represent the uniform basis selection method, the adaptive basis selection method, and the proposed space-filling basis selection method, respectively. 
The time for calculating the smoothing parameter is not included. 
The result for the full basis smoothing spline estimator is omitted here due to the huge computational cost.
The computational time for generating a set of design points, i.e.,  Step 1 in the proposed algorithm,  is not included since the design points can be generated beforehand.
It is observed that the computing time of the proposed method is comparable with that of the other two basis selection methods under all settings. 
Combining such an observation with the result in Fig.~\ref{plot3}, it is concluded that the proposed method can achieve a more accurate prediction, without requiring much more computational time.

\section{Real data example}

The problem of measuring total column ozone has attracted significant attention for decades.  
Ozone depletion facilitates the transmission of ultraviolet radiation (290–400 nm wavelength) through the atmosphere and causes severe damage to DNA and cellular proteins that are involved in biochemical processes, affecting growth and reproduction.
Statistical analysis of total column ozone data has three steps. In the first step, the raw satellite data (level 1) are retrieved by NASA. Subsequently, NASA calibrates and preprocesses the data to generate spatially and temporally irregular total column ozone measurements (level 2). Finally, the level 2 data are processed to yield the level 3 data, which are the daily and spatially regular data product released extensively to the public.

\begin{figure}[ht]
    \begin{center}
        \begin{tabular}{cc}
            \includegraphics[scale = .175]{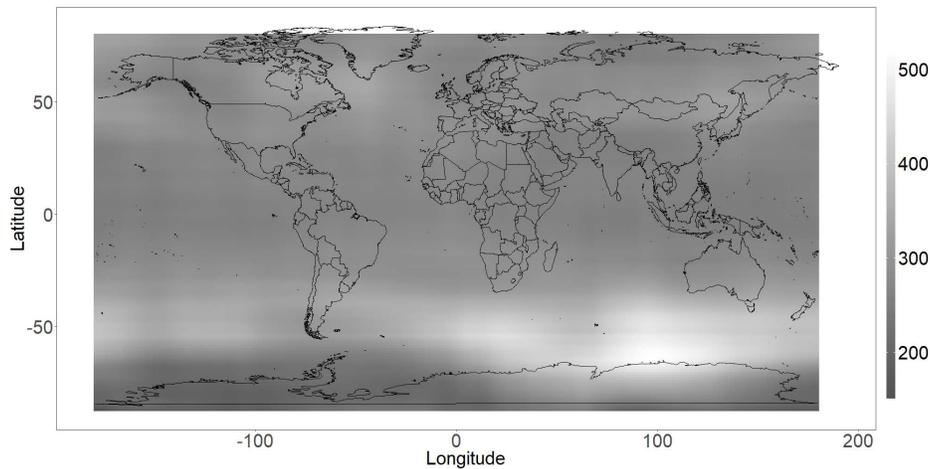}  \\
        \end{tabular}
        \caption{Smoothing spline prediction of total column ozone value for 10/01/1988, in Dobson units}\label{fig_ozone_pred}
    \end{center}
\end{figure}

We fit the nonparametric model 
$y_{ij}=\eta(x_{\langle 1\rangle i},x_{\langle 2\rangle j})+\epsilon_{ij}$ to a level 2 total column ozone dataset ($n$=173,405) compiled by \citet{cressie2008fixed}.
Here, $y_{ij}$ is the level 2 total column ozone measurement at the $i$th longitude, i.e., $x_{\langle 1\rangle i}$, and the $j$th latitude, i.e., $x_{\langle 2\rangle i}$, and $\epsilon_{ij}$ represent the independent and identically distributed random errors. The heat map of the raw data is presented in the Supplementary Material.
The thin-plate smoothing spline is used for the model-fitting, and the proposed method is employed to facilitate the estimation. 
The number of basis functions is set to $q=20n^{2/9}\approx 292$.
The design points employed in the proposed basis selection method are yielded from a Sobol sequence \citep{dutang2013randtoolbox}. 
The heat map of the predicted image on a $1^{\circ}\times 1^{\circ}$ regular grid is presented in Fig. \ref{fig_ozone_pred}. 
It is seen that the total column ozone value decreases dramatically to form the ozone hole over the South Pole, around --$55^{\circ}$ latitudinal zone.

We now report computing times of the model-fitting that are performed on the identical laptop computer for the simulation studies. 
The computational times, in CPU seconds, are presented in parentheses, including basis selection (0.1s), model fitting (129s), and prediction (21s). Further comparison between the proposed method and other basis selection methods on this dataset can be found in the Supplementary Material.

\section{Acknowledgement}
This study was partially supported by the National Science Foundation and the National Institutes of Health.

\bibliography{ref_SBS}
\bibliographystyle{plainnat}

\end{document}